\RequirePackage{fix-cm}
\documentclass[smallextended]{svjour3}

\smartqed

\usepackage{url}
\usepackage{nicefrac}
\usepackage{gensymb}
\usepackage{enumerate}
\usepackage{makecell}
\usepackage[utf8]{inputenc}
\usepackage{graphicx}
\usepackage{caption}
\usepackage{amssymb}
\usepackage{amsmath}

\journalname{Experimental Astronomy}

\begin{document}

\title{Design and construction of a high resolution, portable and low-cost positioner by a star tracking system}
\author{Meysam~Izadmehr, Mehdi~Khakian~Ghomi}
\institute{Meysam~Izadmehr, Mehdi~Khakian~Ghomi \at Department of Energy Engineering and Physics, Amirkabir University of Technology, 15875-4413, Tehran, Iran}

\maketitle

\begin{abstract}
In normal observation procedures, the position of the observer is specified by GPS and celestial positions of an object will be calculated. But in some situations, like small zenith angle FOVs, GPS doesn't work. Therefore in this study, the work is investigated in reverse of usual procedures. Comparison between the local position of a few stars in a typical picture and their reference positions in J2000 is used for determining our location coordinates. Longitude, Latitude and North direction are determined by the procedure. The stars are identified using their unique patterns with neighbor stars. The direction of the sky photography is determined by a designed inclinometer which is harmonious to a camera. For the purpose, it was corrected atmospheric refraction effect, relativistic aberration. It was used half seconds exposure time images to obtain the best results. Therefore this positioning system is used to obtain longitude and latitude with an accuracy of less than 0.503 and 0.816 arc-minutes respectively. Total weight of the system is 4.7 kg, which makes it quite portable.
\end{abstract}

\keywords{star tracker\and positioning system\and determining latitude and longitude\and digital zenith cameras }

\section{Introduction}
Tehran is a metropolitan with about 15 million people at night, therefore there is a great light pollution there. Tehran is on the southern border of Alborz mountain chain which behaves like a high barrier in the northern part of Tehran. So the valleys there (especially with closed southern horizon) are good places to overcome this problem. In these locations, the FOV of the observers is quite limited. Therefore GPS satellites may not be in FOV. Also, in the neighborhood of constructions, GPS coordinates may systematically be distorted. This is the problem of so many astronomical groups who go to these locations. So we tried to investigate the position as an astronomical problem by the stars observed in the limited FOV. One of our goals in this project is to enhance our position accuracy with taking pictures of the sky available to the observer. 

Over recent decades, significant progress has been made in astrogeodetic research and astrometry with the development of digital zenith camera systems\cite{hirt2010}. A zenith camera is an astrogeodetic telescope which is used for the local surveys of the earth's gravity field. University of Hannover in Germany and Geodesy and Geodynamics Laboratory, ETH Zurich of Switzerland, have upgraded the simulated zenith photographic instruments TZK2 and TZK3 with CCD sensor instead of quondam photographic film and developed digital zenith camera systems TZK2-D and DIADEM, respectively \cite{hirt2001,burki2004}. These instruments could be used to observe the apparent motions of stars for the rapid determination of the vertical deflection combined with the geodetic latitude and longitude measured by GPS \cite{wang2014}.

The first star tracker was constructed based on CCD by Salomon in 1976 in JPL\footnote{Jet Propulsion Laboratory} \cite{liebe1993}. Satellites use a star tracker for the attitude determination of satellites; they can measure angles between their local coordinate system and reference coordinate system. For a ground observer, an inclinometer can determine the direction of gravity. As a result, an observer is able to determine its position using the output of the star tracker and the inclinometer. The star tracker takes some photos from the sky; stars of the image are identified, then the rotation matrix  is calculated between the reference coordinate system and its star tracker coordinate system. Since the accuracy of the recorded stars in our image processing is 3 arc-seconds, therefore, we need to consider $i)$ precession, $ii)$ nutation, $iii)$ proper motion, and $iv)$ atmospheric refraction, as well as earth rotation. These corrections are necessary for obtaining longitude, latitude, and north direction with 1 arcminute accuracy. Our procedure consists of three different parts: image processing, determination and recording of camera direction with gravity, and related calculations for finding longitude, latitude, north direction, and their errors.

Section 2 contains the image processing details. Section 3 presents the camera direction calibration. Calculation of positioning and its procedure as the most important part of the work is presented in Section 4. Errors are determined in Section 5. It is presented the results in Section 6 and the concluding remarks are presented in the last section.

\section{Image processing}
The positioning method is based on the night sky photography and its image processing. The first stage is to take some pictures from the night sky using a digital camera. Canon EOS 5D Mark II, 21 MegaPixels (5616$\times$3744) and 6.4$\mu$m pixel size and a skywatcher telescope 120/600 were used for these taking pictures. Skywatcher 120/600 is a refractor telescope with 120mm aperture and 600mm focal length. Usually in each trial it is needed to take about 50 pictures. Canon EOS 5D Mark II and skywatcher 120/600 provide $3.43\degree \times 2.30\degree$ FOV with $2.2$ milliarcseconds resolution. 

To measure the star position in the image,  a point spread function  (PSF) implemented. The used approach, presented by Anderson \& King \cite{anderson2000toward}. This implementation can measure center of a star with the precision of 0.02 pixel. 

The output of this step is equivalent to the unit vectors in "camera coordinate system" and international celestial reference frame (ICRF).

\subsection{Spherical trigonometry approach}
Different algorithms are used to extracting stars in an image \cite{cole2006,jansen2002,samaan2003}. One of these algorithms is based on planar triangle method \cite{samaan2003}. In this work, since it is needed more accurate results, we used spherical triangle geometry instead on planar triangles. In normal photos, border distortion is a common abberate. Since angles of spherical triangles are less sensitive to calibration errors, therefore, this feature used to find the stars \cite{samaan2003}. The advantage of this method over the planar triangle is more independence of this method on the field of view. Wider FOV helps us to find a unique answer for the photographed part on the sky. Another advantage of this method is higher accuracy, which reduces error in calculation of the triangles. The method with higher accuracies help us to find the unique position of the observed of sky by using fewer stars, which reduces the searching time.

To verify the method, Monte Carlo simulations used to produce imaginary stars (brighter than magnitude 9) on the taken pictures ($3.43\degree \times 2.30\degree$ FOV). A false star is a random point which is not pointed to a star; in a real image, it may be a planet or a satellite. In this simulation, false stars add as the brightest stars, so they have maximum effect on star identification. The results of the 1000 simulations are shown in table\ref{table:spherical_triangle_monte_carlo}; Lack of enough stars in the image is the main reason for their incorrect identification.

\begin{minipage}{.9\linewidth}
\def\arraystretch{1.5}
\centering
\bigskip
\label{table:spherical_triangle_monte_carlo} 
\begin{tabular}[t]{ c c c }
\hline \hline
\textbf{Simulate} & \textbf{Number of false stars} & \textbf{Correct results (\%)} \\
\hline
1000 & 0 & 99.2 \\
1000 & 1 & 98.1 \\
1000 & 2 & 96.7  \\
\hline
\end{tabular}\par
\captionof {table}{It has been simulated 1000 simulation for each item. It has added 0, 1 and 2 false stars and it is tried to find the unique position of each photo in the sky}
\end{minipage}

\subsubsection{Creating catalog of spherical triangles}

Since it is needed to identify the coordinates of each star in the sky, The Spherical Triangles Virtual Constellations Catalog (STVCC) has been created. To create it, Tycho-2 catalog is used as the reference data \cite{perryman2010}. Considering the used camera and telescope with 0.5-second exposure time, the stars brighter than magnitude 9 could be detected. This limitation is applied for selection of the stars among Tycho-2 catalog too. Therefore, 124799 stars are selected from Tycho-2 catalog. Since the pictures from the sky should compared with STVCC, therefore, the distance between each two stars should be less than the FOV of the telescope. This limit reduces the number of triangles. At first, size of the sides of each triangle is calculated: 
\begin{equation}\label{eq:3}
\cos a = \cos\delta_1\cos\delta_2\cos(\alpha_1 - \alpha_2)+\sin\delta_1\sin\delta_2
\end{equation}
where $\delta$ is declination and $\alpha$ is right ascension of each star. Angles of a spherical triangle are calculated as follows \cite{smart1977}:
\begin{equation}\label{eq:4}
\cos A=\frac{cos a -\cos b \cos c}{\sin b\sin c}
\end{equation}
Size of the angles and Tycho-2 identifier of the stars are stored in the STVCC. By using STVCC, each sky zone can be identified with only 5 stars with the accuracy of more than \%95, 6 stars \%99, and 7 stars \%99.9.

\subsubsection{Making spherical triangle in camera coordinate system}

\begin{figure} 
\centerline{\includegraphics[width=0.5\textwidth]{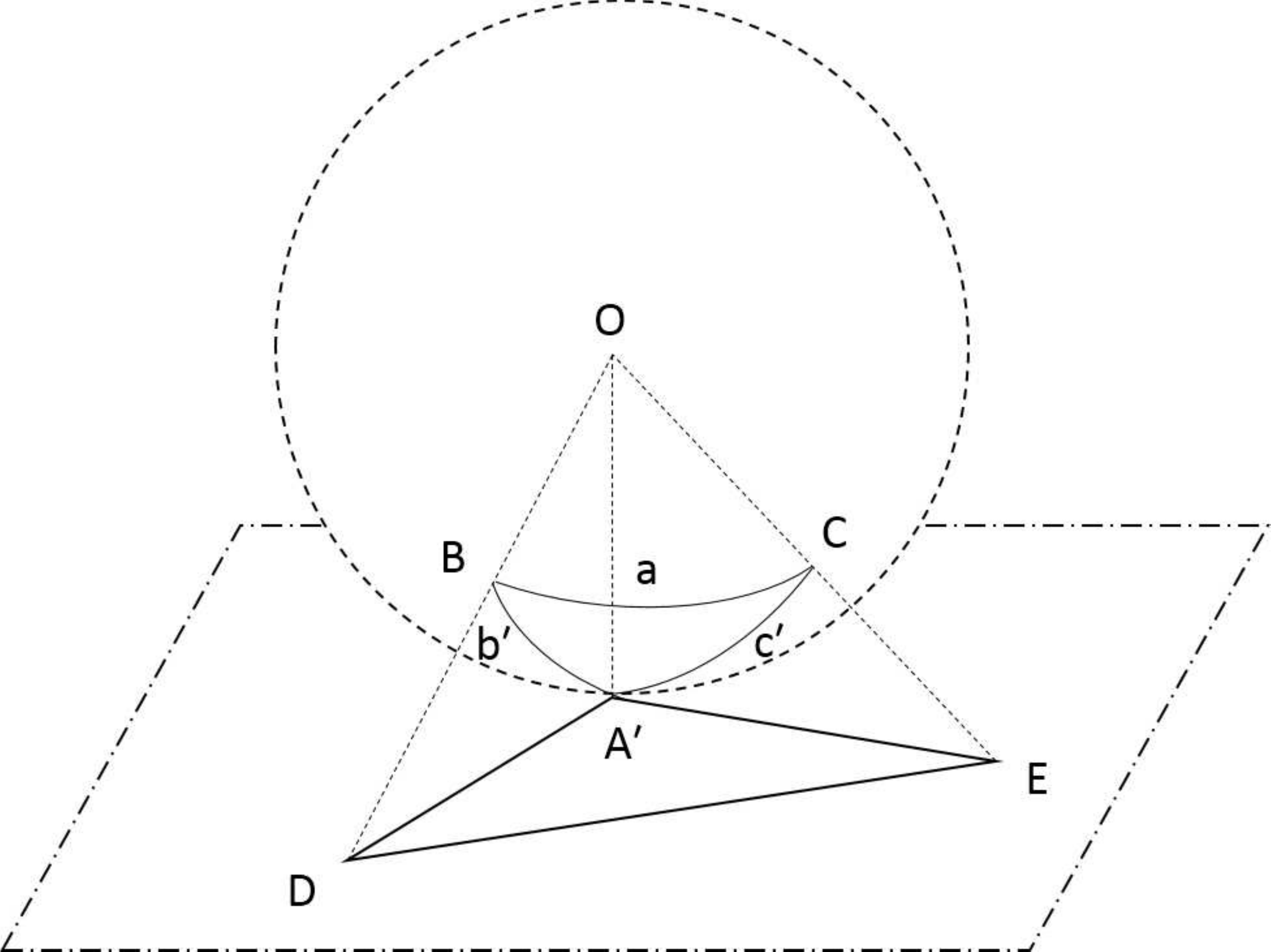}}
 \caption{Celestial sphere and picture frame. $A'$  is the center of the picture, $D$ and $E$ are two arbitrary stars in the picture frame. $B$ and $C$ are their corresponding stars on the celestial sphere
 \label{fig:Celestial_sphere_and_screen}}
\end{figure}

In each spherical triangle $A'BC$ (Figure~\ref{fig:Celestial_sphere_and_screen}) details of the triangle could be calculated as:
\begin{equation}
\tan b'=\frac{DA'}{OA'}
\end{equation}
\begin{equation}
\tan c'=\frac{EA'}{OA'}
\end{equation}
\begin{equation}
\cos A'=\frac{\textbf{\textit{DA}}'\cdot\textbf{\textit{EA}}'}{|\textbf{\textit{DA}}'||\textbf{\textit{EA}}'|}
\end{equation}
Where $OA'$ is the focal length of the telescope, equal to 600mm. the length of each side could be calculated by Eq.~\ref{eq:3}.

It should be noted that, in triangle $A'DE$, two points $D$ and $E$ indicate stars and $A'$ is the center pixel of the image, not a star. Finally, after calculating the of sides of a spherical triangle, the angles are calculated by Eq.~\ref{eq:4}.

There was some assumption in the calculation of the spherical triangle in the camera coordinate system, the center of the CCD is the center of the focal plane, and the CCD is orthogonal to the optical axis of the telescope. Their effects on the accuracy of the system has been tested. Since these effects are much  smaller than the error of the system error, therefore they were eliminated for the calculation.

\subsubsection{Search in STVCC}

To obtain the most accurate position of the taken picture in the sky, it is better to use the most number of recorded stars in the picture. Our instrument is able to identify stars up to magnitude 9, so 124799 stars are selected from Tycho-2. Number of the spherical triangles from these stars is combination of 3 from 124799 which is extremely large. Fortunately, there is a limitation over the problem which reduce the number and it is instrument FOV. In the first stage, the STVCC is created for stars with magnitude up to 7, which contains 951181 spherical triangles with sides smaller than the FOV. Next step is to search for the stars of an image, therefore they are ordered in terms of their brightness. Then, a spherical triangle is created between the first three brightest stars of the image ($S_1, S_2, S_3$) and the spherical triangle is searched in STVCC.  If there is more than one answer for the triangle, the second triangle is created, which includes $S_1, S_2, S_4$ stars. This new triangle is searched in STVCC the same as the first 3 stars. Between the two triangles $S_1S_2S_3$ and $S_1S_2S_4$ there is a common side $S_1S_2$. Therefore the results which do not have two common stars are removed from two lists of the answers. The trend of the triangle creation continues until reaching to a unique answer. 

Since the picture has some aberrations and false stars, therefore, stars brighter than magnitude 9 are used in this FOV. Least square method is used to find the best fit between the picture and Tycho-2 catalog. This procedure is the same as the upper procedure but only for this part of the sky and a new catalog around the identified stars of the picture up to 9 magnitudes.

\subsubsection{Star vectors}

After identification of the stars in the image, the corresponding star vectors are constructed in the reference and the camera coordinate systems. Origin of the camera coordinate system is the central pixel of CCD (point $A'$ in Figure~\ref{fig:Celestial_sphere_and_screen}), X and Y axes directions are on 5616 and 3744 pixels on camera CCD sides respectively. A star vector is a unit vector from the origin of the coordinate system to each star. For each star in the reference coordinate system, the star vector is constructed using the information of its obtained right ascension and declination with the following equation:
\begin{equation}
\textbf{V} = ( \cos \delta \cos \alpha , \cos \delta \sin \alpha, \sin \delta )
\end{equation}
To construct the star vector in the camera coordinate system, the following relations are used:
\begin{equation}
\theta = \arctan(\frac{DA'}{OA'}) \quad,\quad \varphi = \arctan(\frac{(DA')_X}{(DA')_Y})
\end{equation}
\begin{equation}
\textbf{W} =(\cos \varphi \sin \theta,\sin \varphi \sin \theta,\cos \delta)
\end{equation}
$DA'$ and $OA'$ are specified in Figure~\ref{fig:Celestial_sphere_and_screen}. $(DA)_X$ and $(DA')_Y$ are X and Y components of star $D$ from center on the image ($A'$). Table  \ref{table:Angles_of_triangles} shows that the relative error of angular measurements is less than $0.01\%$. For example, it has been calculated 3 angles of 3 real stars HIP104333, HIP104028, and HIP104542 in reference system and camera coordinate system, independently.

\begin{minipage}{0.9\linewidth}
\def\arraystretch{1.5}
\centering
\bigskip
\label{table:Angles_of_triangles} 
\begin{tabular}[t]{c c c c}
\hline \hline
 & STVCC & Image & Relative \\
 & triangle & triangle & error \\
 & (radians) & (radians) &  \\
 \hline
1st angle & 2.27327 & 2.27374 &  $2.07\times 10^{-4}$ \\
2nd angle & 0.634202 & 0.633912 & $4.57\times 10^{-4}$ \\
3rd angle & 0.234213 &  0.234019 & $8.28\times 10^{-4}$ \\
\hline
\end{tabular}\par
\captionof {table}{Angles of triangle constructed in image and STVCC for HIP104333, HIP104028, and HIP104542 stars}
\end{minipage}

\section{Camera-direction calculation}
For the purpose of the positioning, deviation angles of the camera CCD with the horizontal plane should be specified with a high accuracy. Two observers at different longitudes and latitudes are able to observe some common stars in their FOVs. Two observers with different positions and a limited FOV are able to see Polaris and its surrounding stars. The angles between the normal vector of camera CCD and the horizon of the observer could be quite specified. The direction of the camera could be determined using an $inclinometer$. The inclinometer could measure the deviation from the horizontal plane. 
\subsection{Inclinometer and its low-cost supporter electronics}
\begin{figure} 
\centerline{\includegraphics[width=7cm]{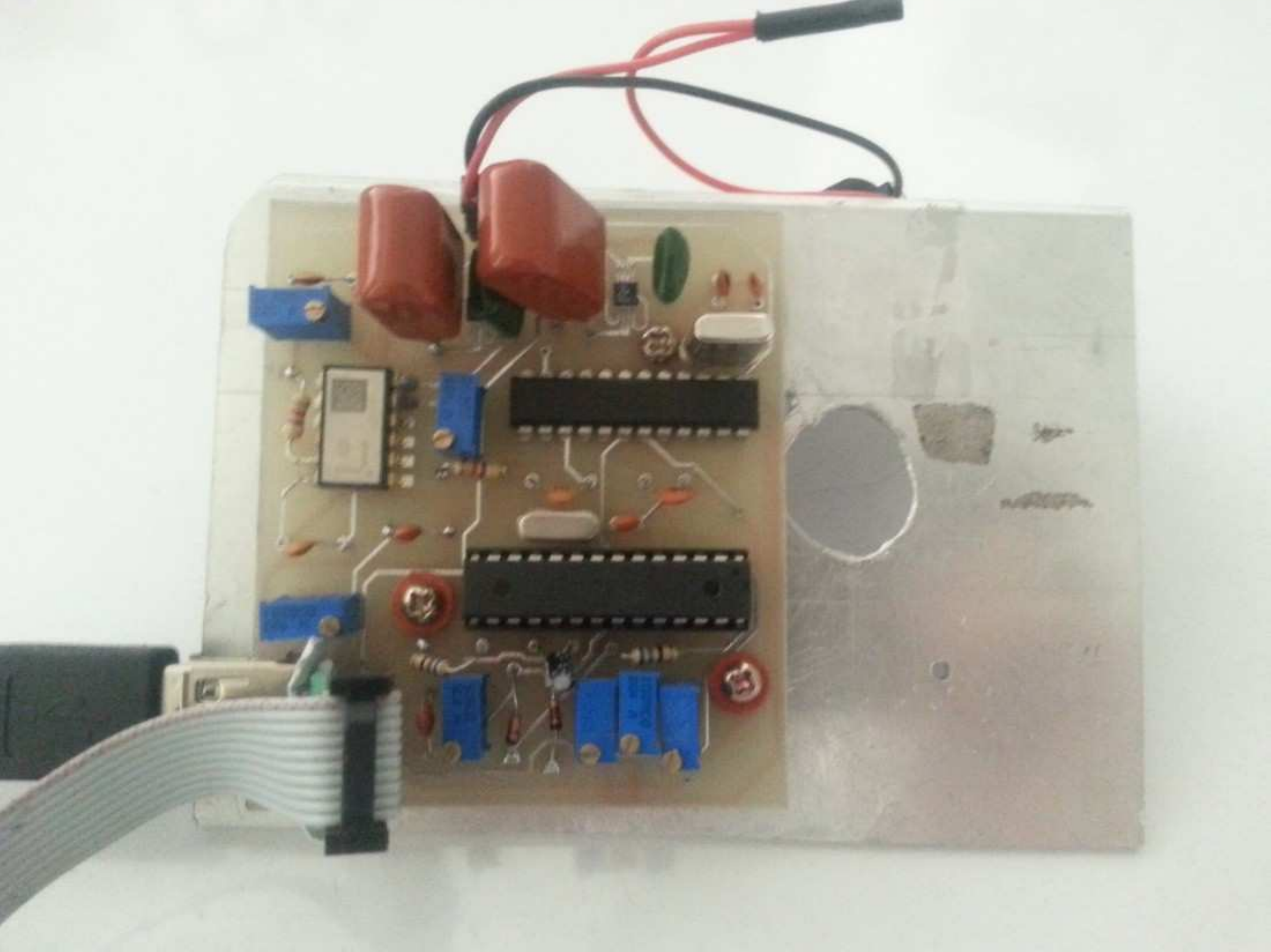}}
 \caption{SCA100T-D01 PCB. The analog output of the SCA100T-D01 converted to the digital output by AD7730. Then Atmega8 reads this digital output and transfers it to PC.
 \label{fig:SCA100T_D01_circuit}}
\end{figure}

Here, a SCA100T-D01 sensor is used as the used inclinometer. SCA100T series have two-axial angle meter ICs based on 3D-MEMS technology, which are used for leveling instruments. The measuring axes of the sensing elements are parallel to the mounting plane and orthogonal to each other. SCA100T-D01 in the interval $\pm 30$ degrees can provide angles with the resolution of 0.0035 degrees with 10 Hz. Angles of two inclinometer axes with the horizontal plane for the analogue output of this sensor are in the relation with voltage:
\begin{equation}
\alpha = \arcsin(\frac{V_{out} - Offset}{Sensitivity})
\end{equation}
$V_{out}$ is the output voltage of the sensor, $Offset$ is output of the sensor for zero angles, i.e. half of input voltage equal to $2.5V$ for $5.0V$ input voltage. $Sensitivity$ indicates the sensitivity of the sensor which is equal to $4V/g$ (4 Volts per gravitational acceleration) for SCA100T-D01. Using SCA100T-D01 outputs, the analog output should be converted into digital output. Because of the high resolution of SCA100T-D01 analog outputs, supporter ADC should have at least 16bits resolution. Microcontroller alone cannot reach this resolution, so AD7730 is used as the external ADC. AD7730 converts the analog output of SCA100T-D01 into the 16-bit digital output. Then, this digital output is transferred to Atmega8 micro-controller. Atmega8 takes the digital output and transfers it to the PC. The constructed inclinometer is shown in Figure~\ref{fig:SCA100T_D01_circuit}.

\section{Positioning procedure}
\begin{figure} 
\centerline{\includegraphics[width=0.5\textwidth]{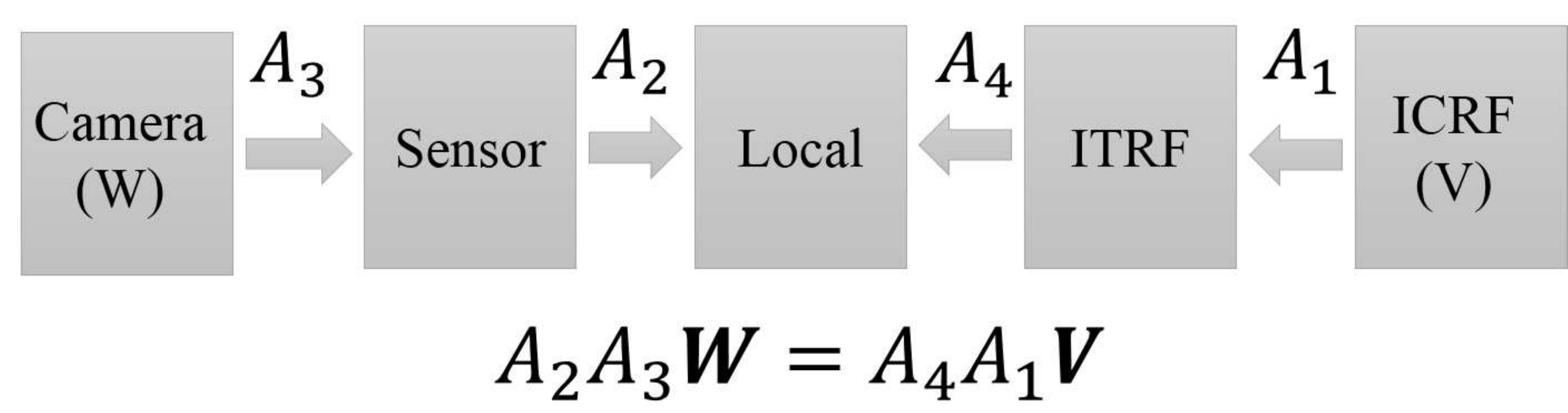}}
 \caption{Schematic design of coordinates conversions. \textbf{W} and \textbf{V} are unit vectors in camera coordinate system and ICRF, respectively. 
 \label{fig:Schematic_design_of_coordinates_conversions}}
\end{figure}

Longitude, latitude and north direction are obtained by the following equation:
\begin{equation}\label{eq:15}
A_2 A_3 \textbf{W} = A_4 A_1 \textbf{V}
\end{equation}
In this equation, $W$ and $V$ are equivalent unit vectors in the camera and reference coordinate system, respectively. $A_i$s are $3\times3$ matrixes.
 \begin{itemize}
\item $A_1$ is the rotational matrix which converts star vectors in ICRF\footnote{International Celestial Reference Frame} into ITRF\footnote{International Terrestrial Reference Frame} 
\item $A_2$ converts sensor coordinate system into the local coordinate system
\item $A_3$ converts camera coordinate system into the sensor coordinate system
\item $A_4$ converts ITRF into the observer's local coordinate system
 \end{itemize}
 Difference between  the observer's local coordinate system and the horizontal coordinate system is their X-axis. It is toward the north in the horizontal coordinate system; however, in the local coordinate system, it is in the observer's horizontal plane. Angle with X-axis of the horizontal coordinate system is one of the outputs from the positioning system.  Astronomical positioning is acquired by the calculation of $A_4$ from Eq.~\ref{eq:15}. It means that the information of the local coordinate system is extracted by the information of reference stars. In Figure~\ref{fig:Schematic_design_of_coordinates_conversions}, coordinate's conversions are schematically shown.

\subsection{Matrix $A_1$}
In Tycho-2 catalog, stars have been recorded by J2000 in terms of the reference coordinate system. The position of the stars relative to each other is almost fixed; most of the proper motions are much less than one arcsec per year\cite{birney2006}. However, they move in the sky because of earth rotation\cite{williams1994,kaplan1989,lowire2007}. Matrix $A_1$ is required for the conversion of ICRF into ITRF. Simulations and frequency fitting are done on the data of earth rotation. New fittings with milliarcsecond precision in each century provide rotation angles. In this work, time series results obtained at IERS\footnote{\textbf{International Earth Rotation} and Reference Systems \textbf{Service}} Conventions (2010) are used for obtaining the rotation matrix\cite{petit2010iers}. 

\subsection{Matrix $A_2$}
Matrix $A_2$ converts sensor coordinate system into the local coordinate system. Outputs of the sensor are the angle of X' and Y' axis of the sensor with the horizon, i.e. $\theta_x'$ and $\theta_y'$, respectively. These angles are the complements of angles with Z-axis of the local system. This matrix is obtained from two rotations: the first one is around Y-axis and the next one is around X-axis. This hypothesis is acceptable because the horizontal coordinate system can be converted into the sensor coordinate system with three rotations first around Z, then Y, and finally X. Matrix $A_4$ contains the rotation around Z. As a result, rotation of  $A_2$ will be as follows:
\begin{equation}
A_2=R_2(b)R_1(a)
\end{equation}
The resulting matrix would be as follows:
\begin{equation}\label{eq:17}
A_2=
\left[ \begin{array}{ccc}
\cos b & \sin a \sin b & \cos a \sin b \\
0 & \cos a & -\sin a \\
-\sin b & \sin a \cos b & \cos a \cos b \end{array} \right]
\end{equation}
Using $\theta_x'$ and $\theta_y'$ results in: 
\begin{equation}
-\sin b =\cos(\frac{\pi}{2}-\theta_{x'})
\end{equation}
\begin{equation}
\sin a \cos b = \cos(\frac{\pi}{2}-\theta_{y'})
\end{equation}
As a result, $a$ and $b$ are equal to: 
\begin{equation}
b= -\theta_{x'}
\end{equation}
\begin{equation}
a=\arcsin\frac{\sin \theta_{y'}}{\cos b}
\end{equation}
Outputs of the sensor are in the interval of -30 to 30 degrees. After determining $a$ and $b$, all the components of matrix $A_2$ can be determined using Eq.~\ref{eq:17}.  

 The relativistic aberration and atmospheric refraction corrections are applied after this matrix multiplied by the vectors \cite{urban2012explanatory}.  

\subsection{Matrix $A_4$}
This rotation matrix is the main part, which is obtained from the following three rotations (Figure~\ref{fig:SCA100T_D01_circuit}):
\begin{equation}
A_4=R_3(c)R_2(\frac{\pi}{2}-\lambda)R_3(\beta)
\end{equation}
$\beta$ is longitude, $\lambda$ is latitude, and $c$ is the angle between X-axis of local coordinate system and direction of the north. In  Eq.~\ref{eq:15}, $A_4$ and $A_3$ are unknown. Since the matrix $A_2$ is between $A_4$ and $A_3$, it is impossible to use attitude determination algorithm for the calculation of the matrix $A_4$. If $A_3$ is known, $A_4$ could be calculated by a deterministic algorithm\cite{shuster2004} or an optimal algorithm\cite{shuster1981,wertz2012}. In this paper, $LMA$\footnote{Levenberg-Marquardt algorithm} is used for calculating matrix $A_4$ \cite{ranganathan2004,lourakis2005}. LMA outperforms simple gradient descent and other conjugate gradient methods is in a wide variety of problems. It is a pseudo-second-order method which means that it works with only function evaluations and gradient information but estimates the Hessian matrix using the sum of outer products of the gradients. $\beta$, $\lambda$, and c are calculated by LMA.
\begin{figure} 
\centerline{\includegraphics[width=0.4\textwidth]{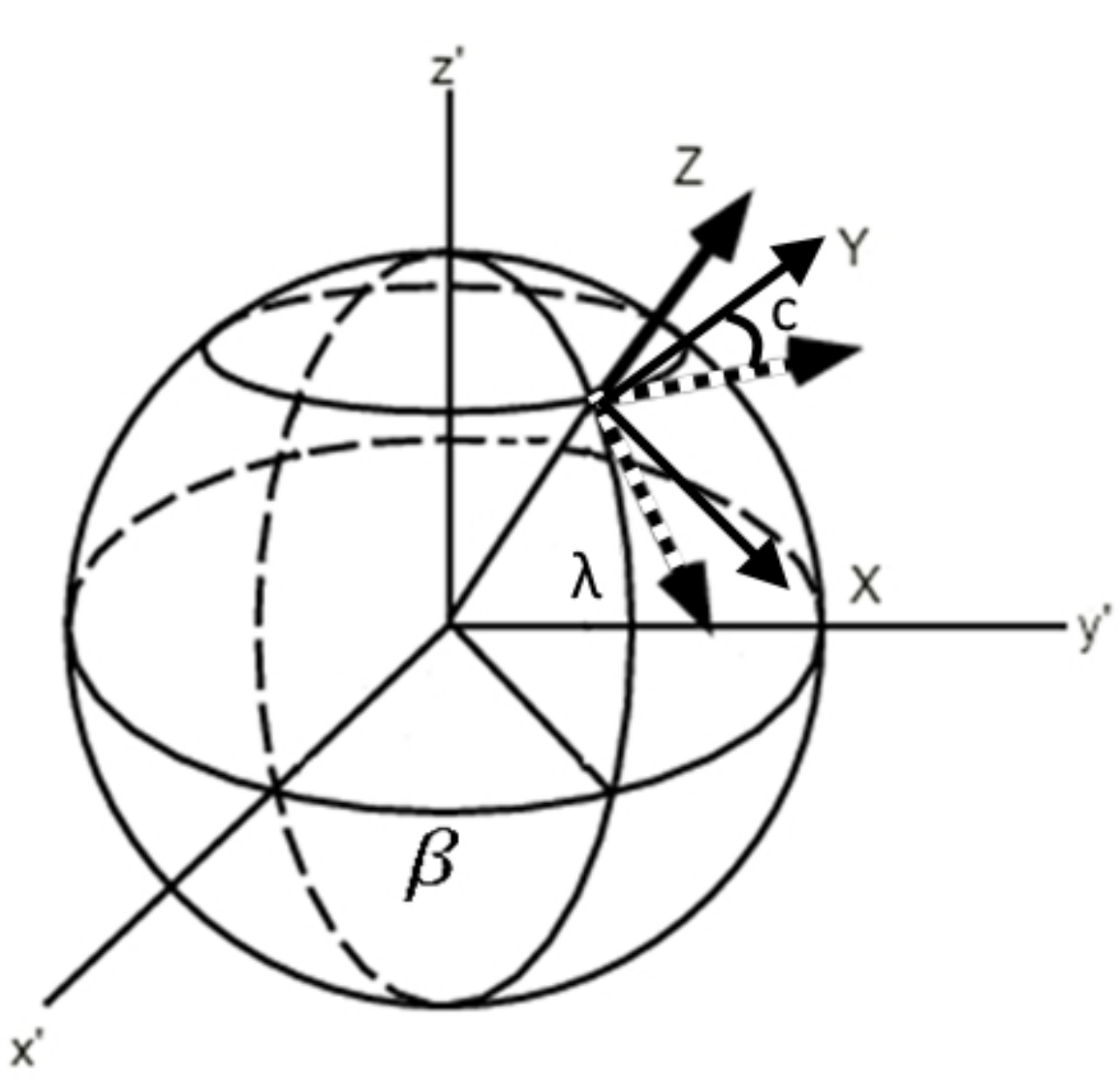}}
 \caption{ (x',y',z') and (X,Y,Z) are ITRF and observer local coordinate system, respectively. 
 \label{fig:ITRF_to_local}}
\end{figure}

\subsection{Matrix $A_3$}
This rotation matrix converts the camera coordinate system into the sensor coordinate system. This rotation matrix is minimized by the calibration procedure. For this calibration procedure, at first, matrix $A_4$ is calculated for a known latitude and longitude and then multiplied by $V$ vectors to obtain vectors in the local coordinate system. Similar to calculating matrix $A_4$, angles of the rotation matrix converting camera coordinate system into the local coordinate system are calculated. Therefore, using these angles, the camera is  rotated to lie on the horizon plane. This rotation is done by using an EQ6 mount which has 0.144 arcsec resolution. The inclinometer is aligned to the plumb line using inclinometer outputs. After fixing the inclinometer to the camera, camera coordinate system is converted into the inclinometer coordinate system with an angle with the Z axis. This angle is fixed, but unknown. By using LMA, there is no need to determine this rotation.

\section{Error investigation}
Using the photography of the night sky and determining the direction of photography, the position of the observer will be determined; however, this positioning is a physical parameter and has some errors. Error sources are:

\subsection{Error in determination of center of the stars}
Considering the specifications of the camera and lens, $2.3^{\circ}$ FOV is located on 3744 pixels. In other words, any pixel covers $\frac{2.3}{3744}=2.21$ arcseconds. This error is for the case where the brightest pixel is selected as a star, but by using the centroiding algorithm, the center of the star can be determined with the accuracy of up to 0.05 pixels \cite{fosu2004}. Also using more than one star, the error of centroiding is decreased.  On the other hand, aberrations increase the error. As a result, the final error of the image processing is the combination of centroiding and aberrations errors, which can be estimated by stability testing. For obtaining this error estimation, the camera is fixed in the observation place. By this fixing, the inclinometer error is removed from the positioning results and stars in the FOV changed by earth rotation.

\begin{figure} 
\centerline{\includegraphics[width=0.7\textwidth]{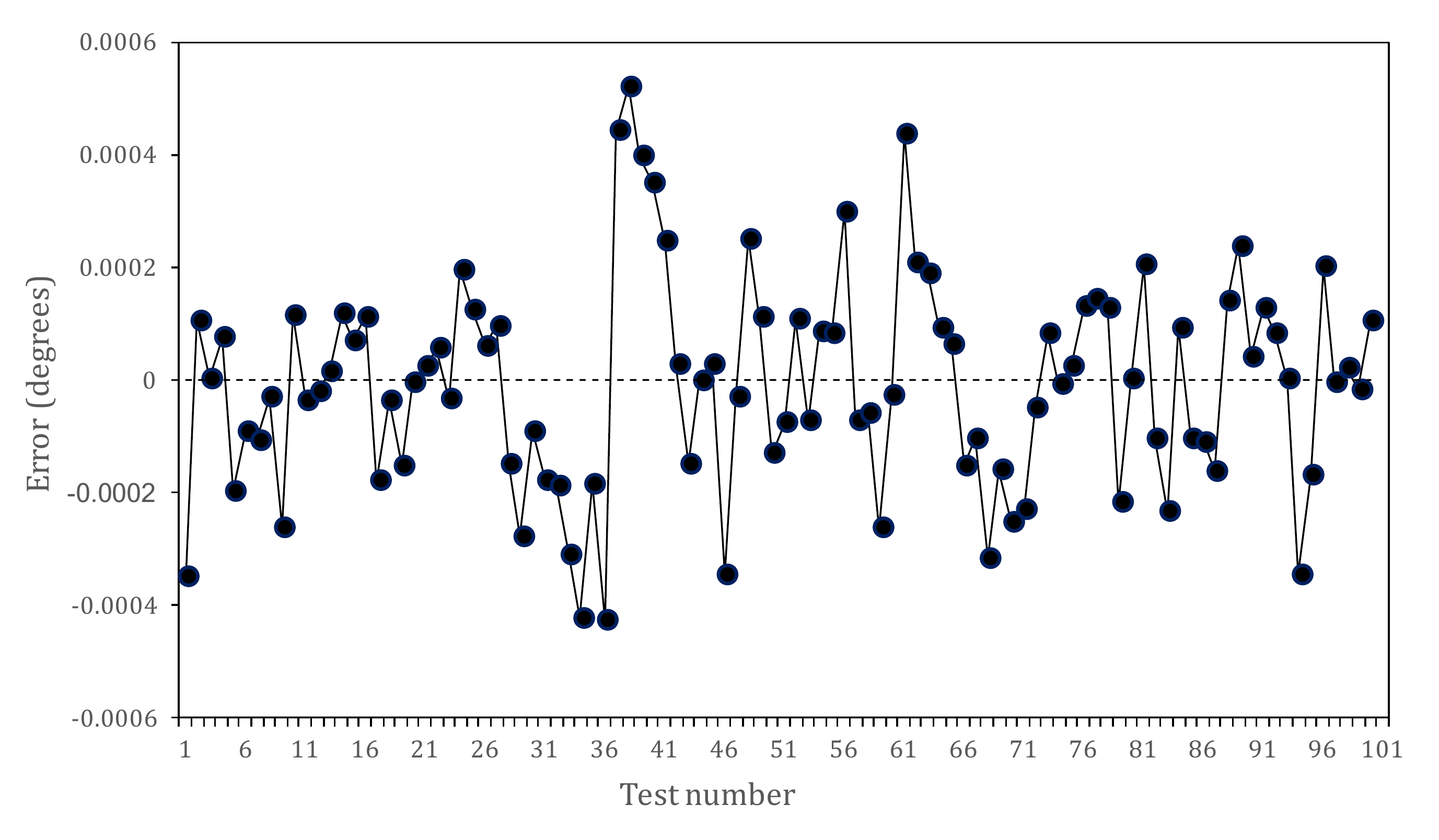}}
 \caption{Latitude error in stability test for 100 images in one direction. Average of the absolute deviation is $0.5256$ arcseconds.
 \label{fig:latitude_error_stability}}
\end{figure}
\begin{figure} 
\centerline{\includegraphics[width=0.7\textwidth]{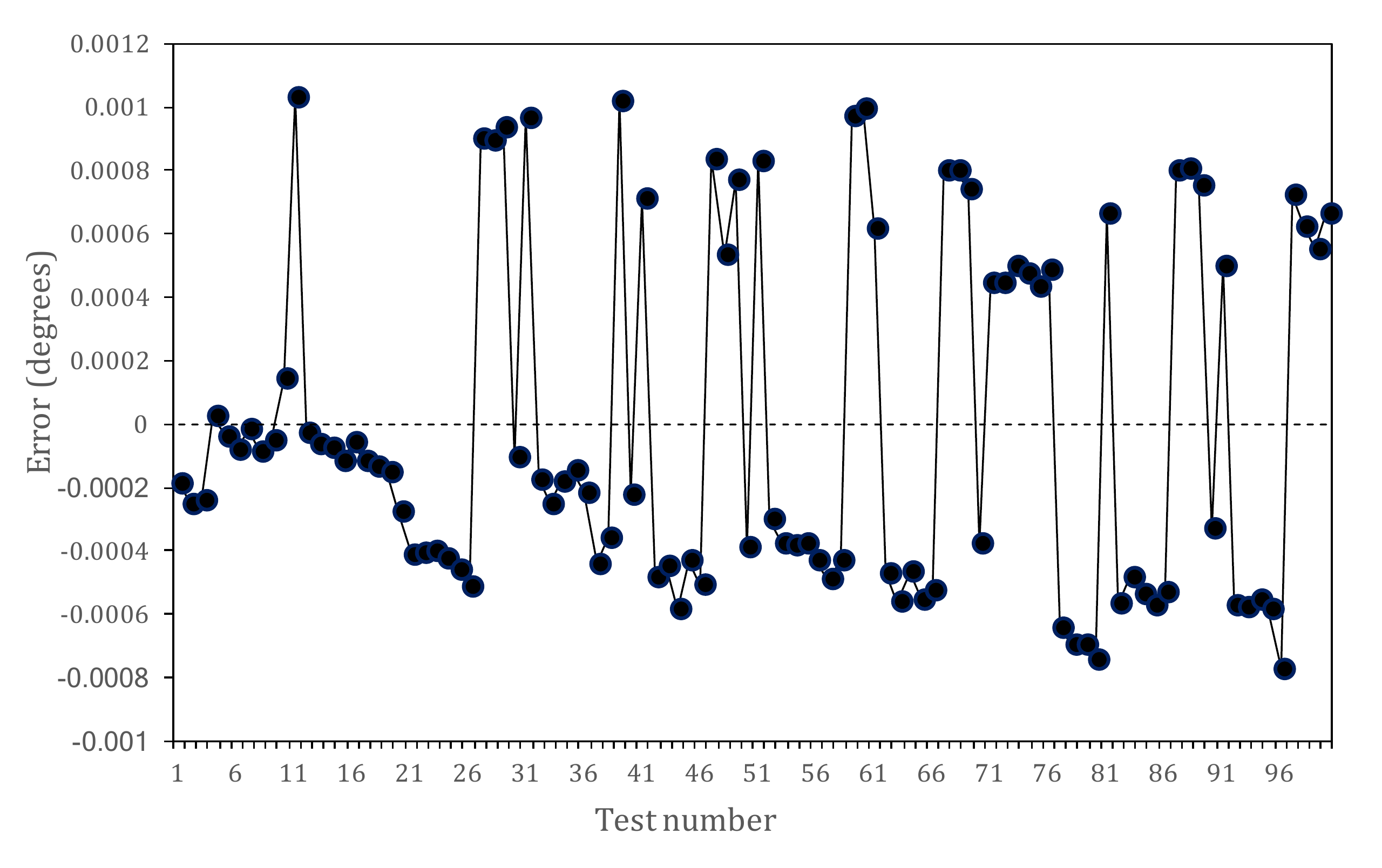}}
 \caption{Longitude error in stability test for 100 images in one direction. Average of the absolute deviation is $1.70$ arcseconds.
 \label{fig:longitude_error_stability}}
\end{figure}
\begin{figure} 
\centerline{\includegraphics[width=0.7\textwidth]{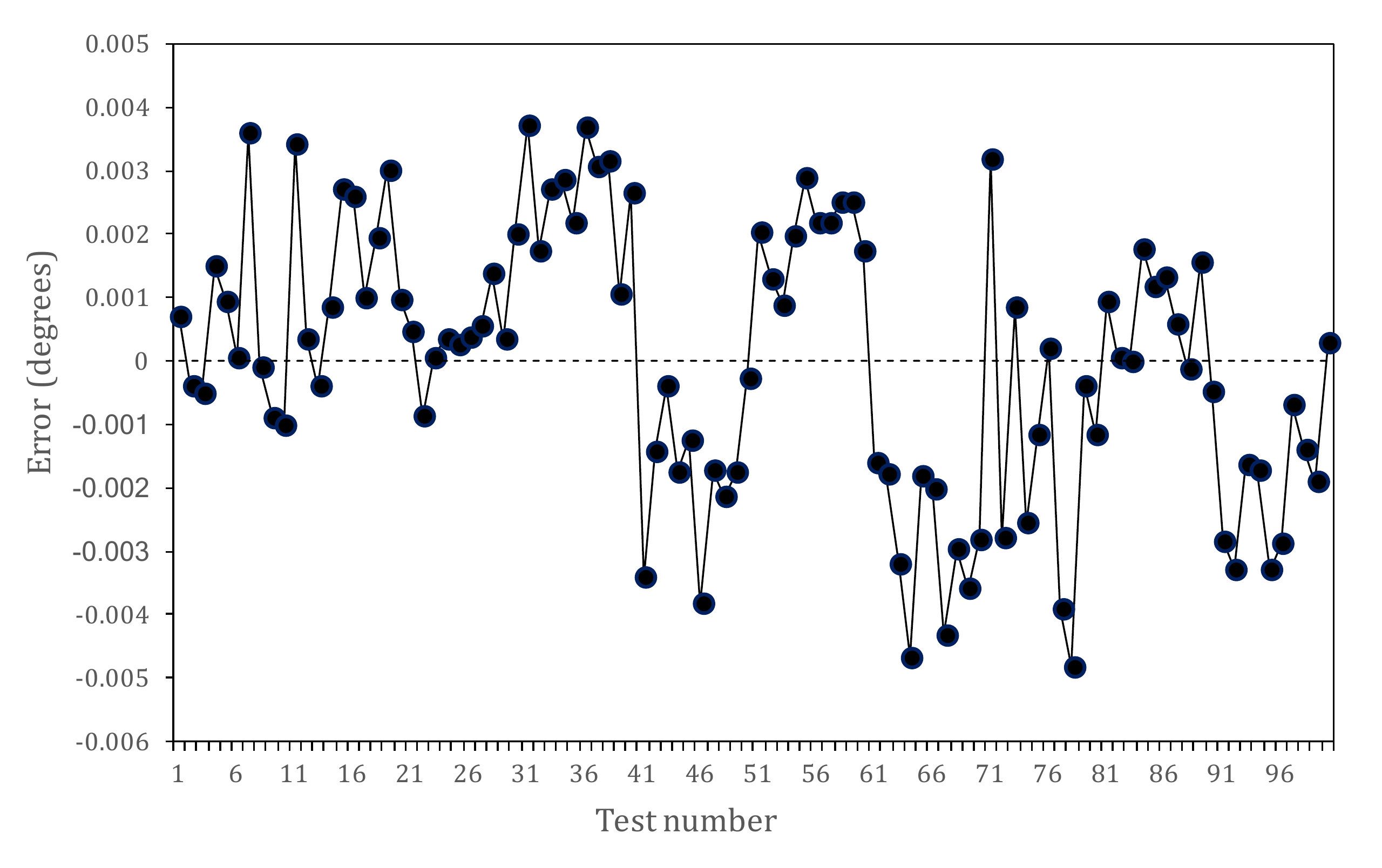}}
 \caption{Error in angle of the north in stability test for 100 images in one direction. Average of the absolute deviation is $6.336$ arcseconds.
 \label{fig:angle_with_the_north_stability}}
\end{figure}

\subsection{Error In determining camera direction}
This error is a systematic error which is due to the inclinometer hardware and it is equal to the resolution of the inclinometer in determining the horizontal plane. The resolution of the used inclinometer is 0.0035 degrees, which is quite enough for this task.

\subsection{Calibration error}
Calibration error can be calculated by comparing the final result with other error sources. it is possible to investigate the other source of error isolated from final test, but it is not possible for error of the calibration.

\section{Results}
\subsection{ Stability Test Results}
The short-term test is done for the error investigation of the position of each star's center in these taken pictures. 100 pictures were taken in one hour in one night. Calculated errors on latitude, longitude and the north direction are shown in Figs.~\ref{fig:latitude_error_stability} to \ref{fig:angle_with_the_north_stability}. In these figures, X-axis is the number of the image and Y axis is the error in degrees.The average of the absolute deviation of these errors in latitude, longitude and angle with the north direction are $0.5256$, $1.70$, and  $6.336$ arcseconds, respectively. 

\subsection{Long duration tests}
\begin{figure} 
\centerline{\includegraphics[width=0.7\textwidth]{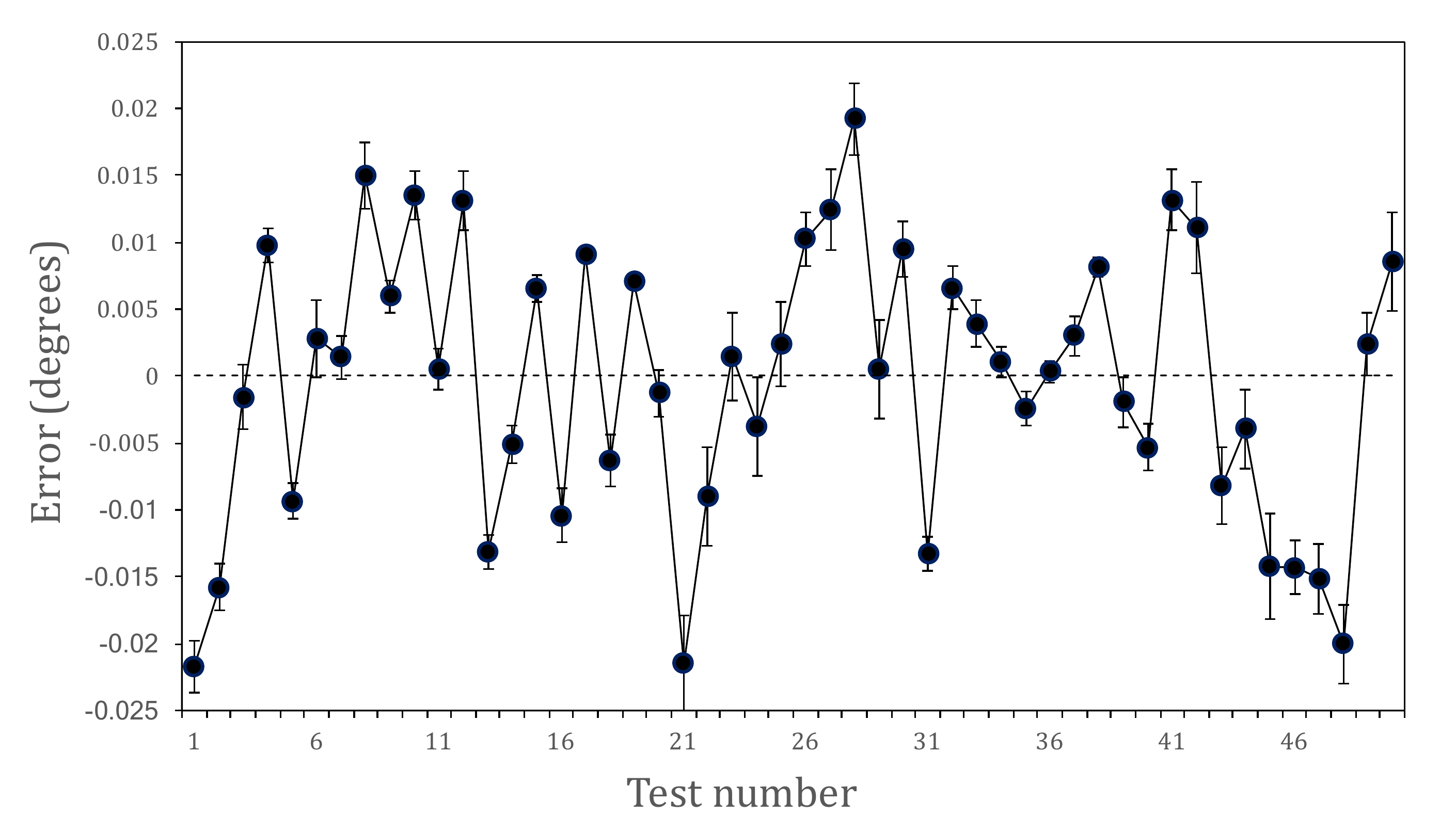}}
 \caption{Latitude error in 50 nights as well as different locations and camera directions. Average absolute deviation of latitude is  $0.5028$ arcminutes. Each point is obtained by at least the average of 100 images.
 \label{fig:latitude_error}}
\end{figure}

\begin{figure} 
\centerline{\includegraphics[width=0.7\textwidth]{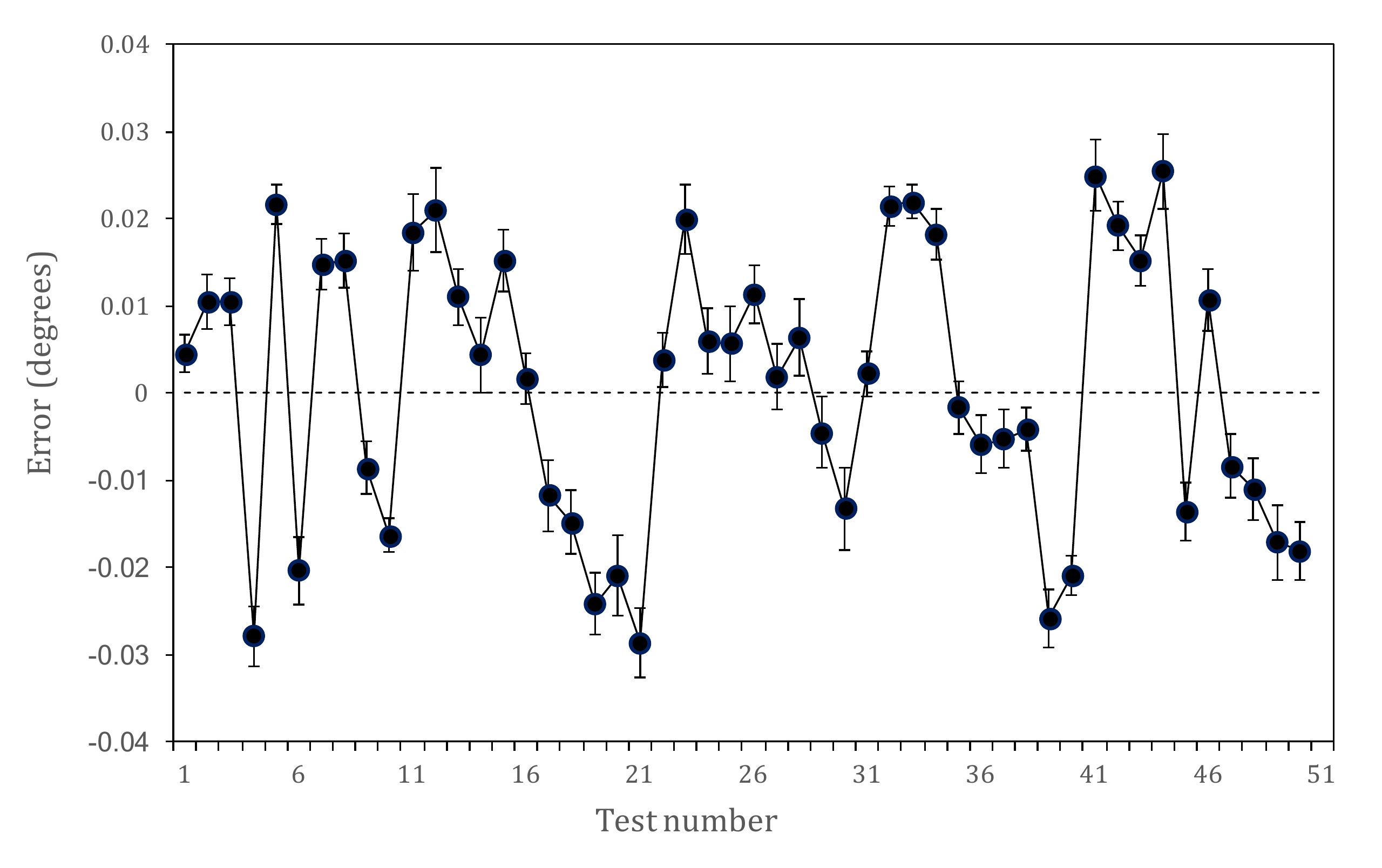}}
 \caption{Longitude error in 50 nights as well as different locations and camera directions. Average absolute deviation of longitude is $0.816$ arcminutes. Each point is obtained by at least the average of 100 images.
 \label{fig:longitude_error}}
\end{figure}

Results of 50 different night tests for latitude and longitude are shown in Figure~\ref{fig:latitude_error} and Figure~\ref{fig:longitude_error}, respectively. In these figures, X axis is number of nights and Y axis is the difference between the calculated and accurate values. Reference longitude and latitude are obtained by GPS values and vertical deflection is from EGM2008\cite{pavlis2012}. During the 50 nights, 50 times tripod is mounted and installed; everything have been reset and photographic situations were quite independent. For the latitude, the mean error is $0.5028$ arcminutes. For the longitude, the mean error is $0.816$ arcminutes. Therefore, the total error is less than 1.5km on the ground. Since  the north direction cannot be determined using another accurate method and the north direction is different for each image, a similar diagram cannot be drawn for the angle with north direction. 

Error of the calibration is $0.28404$ and $0.57762$ arcminutes for latitude and longitude, respectively. This error comes from the calibration procedure, which is explained in section 4.4. Although this error is fixed for the setup, but this matrix multiplied to matrix $A_2$ which is not fixed. Therefore, it is impossible to increase the accuracy by subtracting this average error from calculated latitude and longitude.

\section{Conclusion}
By using stars for positioning, it is possible to determine latitude, longitude, and angle from the true north. One advantage of this type of positioning is no need to communicate with the ground stations and satellites. As \cite{hirt2001} and \cite{burki2004} used this method for determining the deflection of vertical axis; this method can be used as a supplement or cross-checking method. Although the accuracy of determining the latitude and longitude by this positioning system is less than GPS, it can measure the angle with true north accurately too. Also, it works in the places with the small FOV.  The average  positioning time is 35 sec and the weight of our system is 4.7kg. 

  Positioning errors are due to the image processing, inclinometer, and calibration. The values of the image-processing error and the inclinometer error are known in the results section. Comparison of the final error with error to image processing and inclinometer error clarifies that the greatest error in the results comes from the calibration. As a result, the first step in reducing the error is the improvement of the calibration method. Afterwards, changing inclinometer can increase the accuracy of positioning. By using better calibration method error can reduce to 0.0035 degrees, the error of the inclinometer. After reaching the to the error of the inclinometer, changing the inclinometer needed to reduce error. There are many inclinometers with an error less than 1 arcsecond. for example, Jewell tilt-meter model 701-2A. In this error range, shutter time needed to reduced and for that telescope with larger aperture should use. By this changes the error of 1 arcsecond can be reached, which means 30m error in the positioning. 

\nocite{*}
\bibliographystyle{spmpsci}
\bibliography{Local_Positioning_System_Using_Star_Tracker}

\end{document}